\documentclass[aps,prc,superscriptaddress,showpacs]{revtex4}
\usepackage[dvips]{graphicx}
\usepackage{amsmath,amssymb}

\newcommand{\vx}{\mbox{\boldmath $x$}}

\newcommand{\vz}{\mbox{\boldmath $z$}}

\newcommand{\vp}{\mbox{\boldmath $p$}}

\begin{document}

\title{Resonant Monopole Oscillation in the Bose-Fermi Mixed
System}\thanks{Proceedings of the 13th International Lazer Physics
Workshop, Trieste, July 12 $-$ 16, 2004 }

\author{Tomoyuki Maruyama\footnote{e-mail: tomo@brs.nihon-u.ac.jp}}
\affiliation{
Institute for Nuclear Theory,
University of Washington,
Seattle, Washington 98195, USA}
\affiliation{
College of Bioresource Sciences,
Nihon University,
Fujisawa 252-8510, Japan }
\author{Hiroyuki Yabu}
\author{Toru Suzuki}
\affiliation{Department of Physics, Tokyo Metropolitan University, 1-1
Minami-Ohsawa, Hachioji, Tokyo 192-0397, Japan}

\begin{abstract}
We study the monopole oscillation in the bose-fermi mixed condensed
system by performing the time-dependent Gross-Pitaevskii  and  Vlasov
equations.
We study
the resonant oscillation where the intrinsic frequencies of
boson and fermion oscillations are same.
\end{abstract}



\maketitle


Over the last few years
there has been  significant progress in the production of 
ultracold gases, which realize the Bose-Einstein condensates (BEC) \cite{BECrv,Dalfovo,bec}, 
degenerate atomic fermi gases\cite{ferG}  and bose-fermi mixtures 
\cite{BferM}.
These systems offer great promise for studies of new, interesting-driven
phenomena.

Collective oscillations are some of the most prominent phenomena 
common to a variety of many-body system.
Theoretical studies on the collective motion in bose-fermi 
mixtures were performed
in the normal phase with the sum-rule approach \cite{miyakawa} and
the random-phase approximation (RPA) \cite{sogo}.

In Ref. \cite{tomoM} we constructed a transport mode for the bose-fermi
mixing system by combining the time-dependent Gross-Pitaevskii(GP) equation
and the Vlasov equation,
and studied the monopole oscillation
by solving the time-evolution of the system directly.
Then we found a rapid damping in the fermion oscillation 
and a blurring of the beat in the boson oscillation. 
We considered the system where the boson number $N_b$ is 
much larger than the fermion number $N_f$ ($N_b \gg N_f$),
which optimizes the overlap of bosons and fermions.
In such systems
the boson intrinsic frequency is $\sqrt{5}$ which has been 
shown in the system including the condensed bosons only \cite{Dalfovo}.
On the other hand the fermion intrinsic frequency becomes larger as
the attractive boson-fermion coupling increases.
This is suggestive that the fermion intrinsic frequency 
becomes equivalent to the boson one for a certain value of the boson-fermion 
coupling.

In this work we choose the value of the boson-fermion coupling
 where the boson and fermion intrinsic frequencies are same.
Then we investigate resonant behavior of
the monopole oscillation by performing the numerical simulation.

\bigskip

Here we briefly explain our formalism.
First we define the Hamiltonian for boson-fermion coexistent system
as follows.
\begin{equation}
H = H_B + H_F + H_{BF}
\end{equation}
with
\begin{eqnarray}
H_B &=& \int d^3 x \{ - \frac{1}{2} \phi^{\dagger}(\vx) \nabla^2 \phi(\vx)
+ \frac{1}{2} \vx^2 \phi^{\dagger}(\vx) \phi(\vx)
+ \frac{g_B}{2} (\phi^{\dagger} \phi)^2 \} ,
\\
H_F &=& \int d^3 x \{ - \frac{\hbar^2}{2m_f} \psi^{\dagger} \nabla^2 \psi
+ \frac{1}{2} m_f \omega_f^2 \vx^2 \psi^{\dagger} \psi \} ,
\\
H_{BF} &=& h_{BF} \int d^3 x \{ \phi^{\dagger} \phi \psi^{\dagger} \psi \} ,
\end{eqnarray}
where $\phi$ and $\psi$ are boson and fermion fields, respectively.
Fermion mass $m_f$ and the trapped frequency $\omega_f$ are normalized
with the boson mass $M_B$ and the boson trapped frequency $\Omega_B$, respectively.
The coordinates are normalized by $\xi_B = (\hbar / M_B \Omega _B)^{1/2}$.
The coupling constants $g_B$ and $h_{BF}$ are given as
\begin{eqnarray}
g_B &=&  4 \pi a_{BB} \xi_B^{-1},  \\
h_{BF} &=&  2 \pi a_{BF} \xi_B^{-1} (1 + m_f^{-1}),
\end{eqnarray} 
where $a_{BB}$ and  $a_{BF}$
are the scattering lengths between two bosons
and between the boson and the fermion, respectively.

The total wave function ${|\Phi(\tau)>}$ has  $N_c$ condensed bosons, whose
wave function $\phi_c$ is defined as
\begin{equation}
\phi_c (\vx,\tau) = <\Phi|\phi_B(\vx,\tau)|\Phi>,
\end{equation}
where $\tau$ is the time coordinate normalized with $\Omega_B^{-1}$.
In this work the wave function $\phi_c$ is expanded 
with the harmonic oscillator wave function $u_n(\vx)$ as
\begin{eqnarray}
&&\phi_c (\vx,\tau) = \sum_{n=0}^{N_{base}} A_n e^{i \theta_n} u_n(b\vx^2) 
e^{-\frac{i}{2} \nu\vx^2} , 
\label{c-tran}
\end{eqnarray}
where $N_{base}+1$ is the number of the harmonic oscillator bases.
We define the Lagrangian with the collective coordinates as
\begin{equation}
L(A_n, \theta_n, b, \nu) = 
< \Phi(\tau)| \{ i \frac{\partial}{\partial \tau} - H \} | \Phi (\tau) >.
\end{equation}
Instead of solving the time-dependent GP equation directly,
we take $A_n, \theta_n, b, \nu$ as time-dependent variables 
and solve the Euler-Lagrange equations with respect to
these variables. 

The many fermion system can be described with the Thomas-Fermi
approximation, which is given in the limit $\hbar \rightarrow 0$.
Here we define the phase-space distribution function  as
\begin{equation}
f(\vx,\vp,\tau) =
\int {d^3 z} <\Phi| \psi(\vx+\frac{1}{2}{\vz},\tau)
\psi^{\dagger}(\vx-\frac{1}{2}{\vz},\tau) |\Phi>
 e^{-i \vp \vz} .
\end{equation}
In this classical limit
this phase-space distribution function satisfies
the following Vlasov equation \cite{KB}:
\begin{equation}
\frac{d}{d \tau} f(\vx,\vp;\tau) =
\left\{ \frac{\partial}{\partial \tau} + \frac{\vp}{m_f}{\nabla_x} -
 [\nabla_x U_F(x)][\nabla_p] \right\} f(\vx,\vp;\tau) = 0 .
\label{Vlasov}
\end{equation}
with
\begin{equation}
U_F (\vx) = \frac{1}{2} m_f \omega^2_f \vx^2 + h_{BF} \rho_B (\vx) ,
\label{uF}
\end{equation}
In the actual calculation we solve the above Vlasov equation (\ref{Vlasov}) 
with the test particle method \cite{TP}.

Thus we can get the time-evolution of 
the condensed bosons and the fermions
by solving the above time-dependent GP equation
and the Vlasov equation.

\bigskip

We calculate the monopole oscillation
of the system $^{39}$K $-$ $^{40}$K;
the number of bosons ($^{39}$K) and fermions ($^{40}$K)
are taken to be 100,000 and 1,000, respectively.
The trapped frequencies are taken to be $\Omega_B = 100$ (Hz) and 
$\omega_f = 1$.
The boson-boson interaction parameters is set to 
$g_B = 1.34 \times 10^{-2}$   which corresponds to 
$a_{BB} = 4.22$(nm) \cite{miyakawa};
in addition we vary the boson-fermion interaction parameter $h_{BF}$
which is measured in the unit of
$h_0 = 7.82 \times 10^{-3}$ corresponding to 
$a_{BF} = 2.51$(nm) \cite{miyakawa}.

\begin{figure*}
\includegraphics[scale=0.7,angle=270]{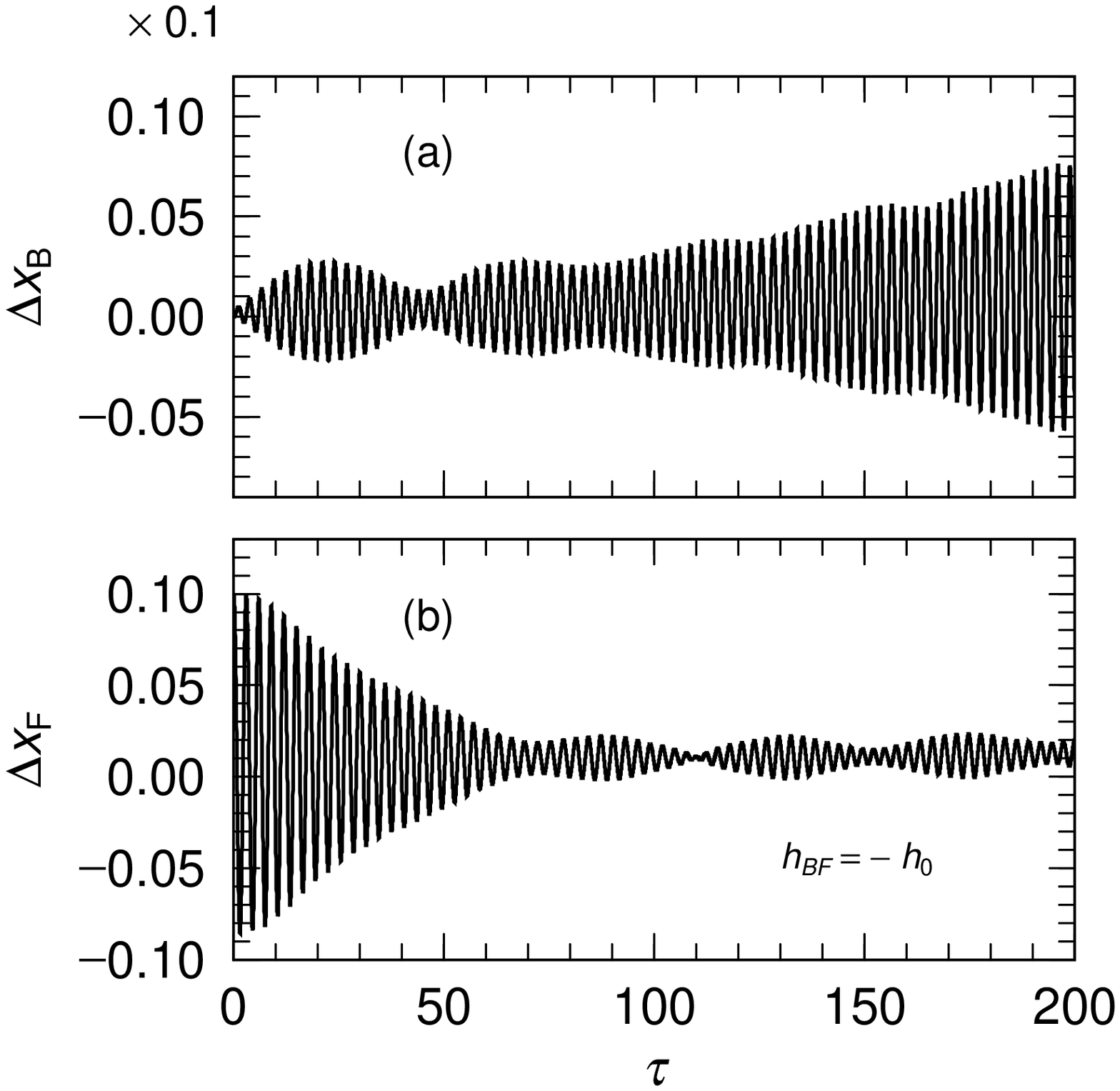}
\caption{
Time evolution of $\Delta x$ for boson (a) 
and fermion (b) at the initial condition that $s_b = 1.0$
and $s_f = 1.1$ with the boson-fermion coupling $h_{BF}=-h_0$.}
\label{rmRx1}
\end{figure*}

Using the root-mean-square radius (RMSR) $R$,  we here define 
\begin{equation}
\Delta x (\tau) = R (\tau) /{R}_0 - 1,
\end{equation}
where  $R_0$~ is the RMSR of the ground state, and $\tau$ is the time
variable normalized by $\Omega_B^{-1}$.
In Fig.~\ref{rmRx1} we show  the time-dependence of this observable for bosons
($\Delta x_B$) and fermions ($\Delta x_F$) with $h_{BF} = - h_0$.
In this calculation the initial condition is taken to be
$\Delta x_B = 0$ and  $\Delta x_F = 0.1$ at $\tau = 0$.
We see a fast damping in the fermion oscillation as shown in Ref.\cite{tomoM}.
However the the boson oscillation grows longer and,
its amplitude becomes larger with $h_{BF} = - h_0$
than  when the boson-fermion coupling is repulsive, $h_{BF} > 0$.

\begin{figure*}
\includegraphics[scale=0.7,angle=270]{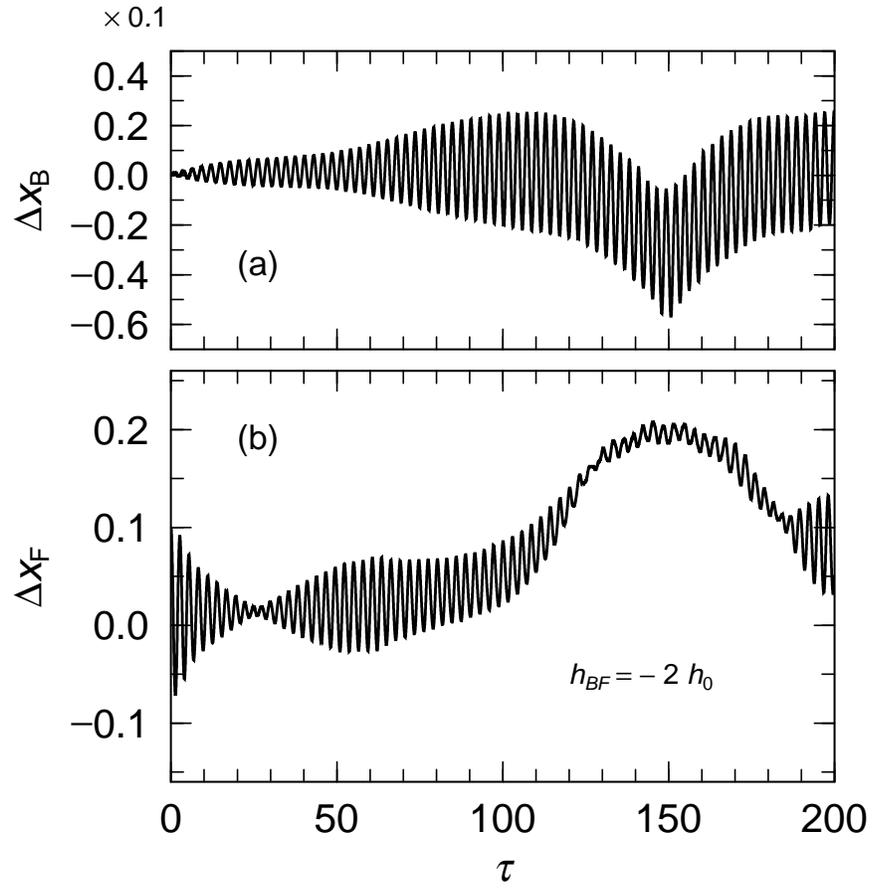}
\medskip
\caption{\small 
Time evolution of $\Delta x$ for boson (a) 
and fermion (b)  with the boson-fermion coupling $h_{BF}=-2h_0$.
The other conditions are same as those in Fig.~\ref{rmRx1} }
\label{rmRx2}
\end{figure*}

When varying the boson-fermion coupling, the situation becomes different.
In Fig.~\ref{rmRx2} we plot the results  with $h_{BF} = - 2h_0$.
The oscillation behaviors are quite different from those in Fig.~\ref{rmRx1}.
We can see the extension and shrink of the fermion density.

\begin{figure}
\includegraphics[scale=0.7]{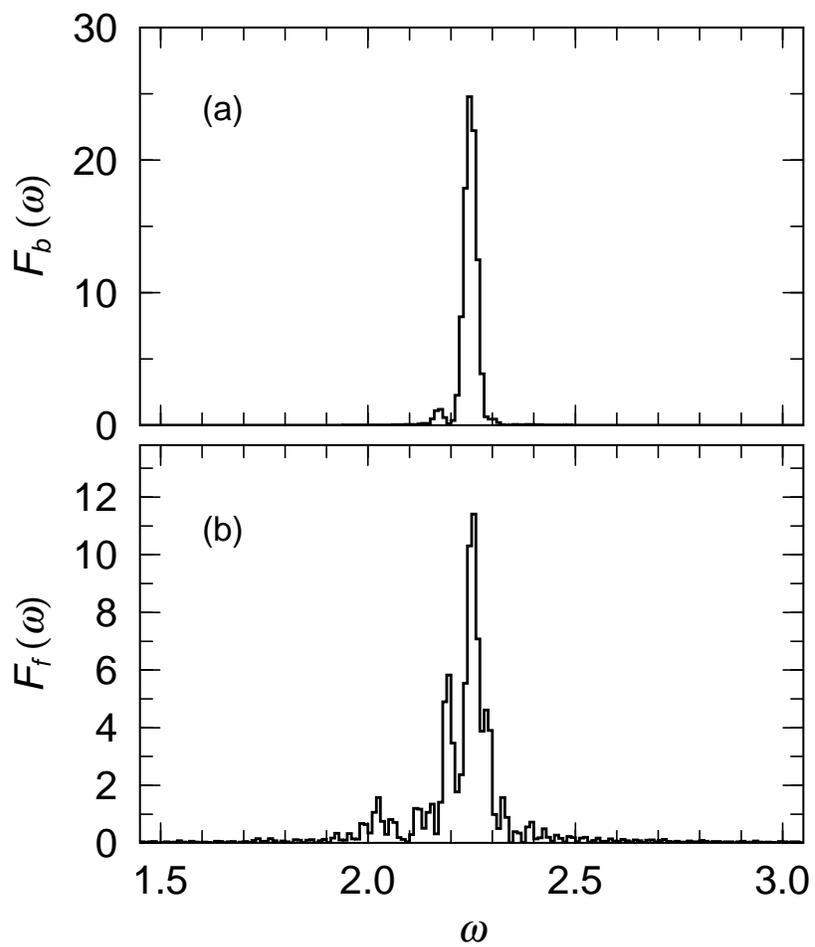}
\medskip
\caption{\small
Spectra of the boson (a) and fermion (b) oscillations
shown in Fig.~\ref{rmRx2}}
\label{spect}
\end{figure}

In order to study this phenomenon, further, we examine the spectrum 
which is obtained by the Fourier transformation
\begin{equation}
F(\omega) = {A_C} \left| \int d \tau R^2(\tau) e^{i \omega \tau} \right|^2,
\end{equation}
where $A_C$ is a normalization factor. 
In Fig.~\ref{spect} we give the spectra for boson (a) and fermion (b)
oscillations obtained by integrating over
the regions $0 < \tau < 200$.
We can see that these two spectra have peaks at the same frequencies.

The oscillation process is explained as follows.
At first the fermion oscillation starts and trigger 
the boson oscillation.
The boson oscillation grows  while
the fermion oscillation very rapidly damps.
Once the fermion oscillation is stopped and grows up, again.
The boson oscillation has a large amplitude at that time and
scatter many fermions to the outside region.
The boson density is distributed in a smaller region than the fermion density.
Outside of the boson density
the fermions feel only  the harmonic oscillator potential, 
which is weaker than attractive force inside the boson density region.
In addition the scattered fermions lose the kinetic energy
and remain there rather long time.
So fermions move from the inside region to the outside region
in the expansion time $30 \gtrsim \tau \gtrsim 150$.

Later fermions far from the boson region return to the central region, 
but some fermions moving out from the boson region 
continue to populate outside the boson region because
a test particle in the harmonic oscillator potential moves on closed orbit.
which does not cross in the boson region.
The system approaches to being equilibrium arounr $\tau \approx 200$.
Hence the averaged radius of fermion density remains to be 10 \% 
larger than that at the ground state    

\begin{figure}
\includegraphics[scale=0.7,angle=270]{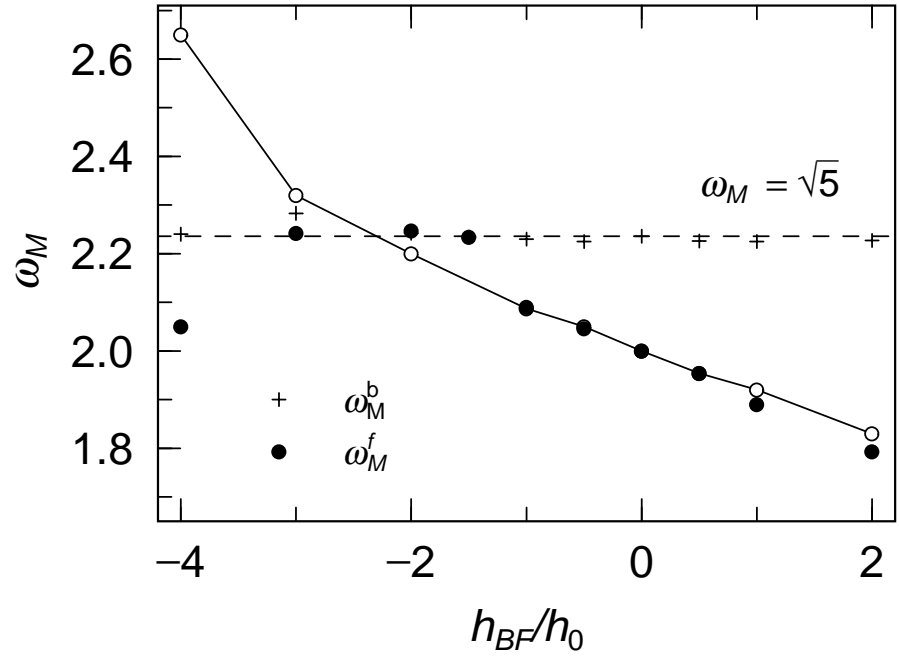}
\medskip
\caption{\small
Intrinsic frequencies of boson and fermion oscillations 
versus the boson-fermion coupling.
Crosses and full squares represent the intrinsic frequencies of boson and
fermion oscillations, respectively.
The open circles, which are connected with the solid line, 
indicate the fermion intrinsic frequency with the boson motion frozen.
The dashed line denotes $\omega_M = \sqrt{5}$.
}
\label{freq}
\end{figure}

In Fig.~\ref{freq} we show the intrinsic frequencies of the boson, $\omega_M^b$, 
(crosses) and fermion, $\omega_M^f$, (full squares) oscillations
as functions of the boson-fermion coupling $h_{BF}$ normalized with $h_0$.  
For reference we also plot the fermion intrinsic frequencies with
the open circles which are connected by the solid line,
and $\omega_M^b = \sqrt{5}$,  which has been 
shown in the system only including the condensed bosons  \cite{Dalfovo},
with the dashed line.
The boson intrinsic mode is almost independent of fermions because  $N_b \gg N_f$.

When the boson-fermion coupling is $h_{BF} \gtrsim - h_0$
the intrinsic frequencies of the fermionic oscillation are almost
equal to the frequency under the boson motion frozen, and  
their values are estimated to be \cite{tomoM}
\begin{equation}
\omega^{M}_f 
\approx 2  ( \omega + h_{BF} \frac{ \partial^2 \rho_B}{\partial x^2} ).
\end{equation}    
The frequencies are determined by the boson density distribution,
and these fermion oscillations are volume vibrations.

When the boson-fermion coupling is $-3 h_0 \lesssim h_{BF} \lesssim -1.5 h_0$,
however,
$\omega_M^f \approx \omega_M^b$, and the resonant oscillation occur.
This oscillation behavior is similar in the interaction region,
though $\Delta x_F$ reaches its maximum  earlier as 
the interaction becomes more strongly attractive.

When the boson-fermion coupling becomes larger, $h_{BF} \lesssim -4 h_0$,
the fermion intrinsic frequency becomes $\omega_M^f \approx 2.1$,
which is close to that only in the harmonic oscillator potential,
$\omega_M^f = 2$.
In strong attractive interactions the fermion motion loses collectivity,
and RMSR of fermion density is affected only by the motion of fermions
outside the boson density region.
Namely the fermion oscillation is surmised to become 
a surface vibration.

In summary we study the collective monopole oscillation in the bose-fermi
mixed condensed system when $h_{BF} = -2 h_0$.
In this coupling the intrinsic frequency of the fermion oscillation, 
$\omega_M^f$, is
almost equal to with the intrinsic frequency of the boson oscillation, 
$\omega_M^b$, and we observed resonant behavior.
In this resonant oscillation the amplitude of the boson oscillation
becomes quite large even if the initial amplitude is taken to be zero.
Furthermore  the fermion density is largely extended 
because some fermions
must move out from the boson density region to achieve the mixed
system equilibrium.

Furthermore we find that the fermion oscillation becomes a surface vibration
when the boson-fermion coupling is more strongly attractive
 $h_{BF} \lesssim -4 h_0$,
than that in the resonant region,  $-3 h_0 \lesssim h_{BF} \lesssim -1.5 h_0$.
While fermion oscillation exhibits the volume vibration 
when the coupling is attractive weaker or
repulsive,  $h_{BF} \gtrsim -1 h_0$. 
In future we would like to examine this consideration more detailed.

\newpage

\end{document}